# Density Enhanced Diffusion of Dipolar Excitons within a One-Dimensional Channel

X. P. Vögele,[1] D. Schuh,[2] W. Wegscheider,[2] J. P. Kotthaus,[1] A. W. Holleitner[1,3,*]

[1] Fakultät für Physik and Center for Nanoscience, Ludwig-Maximilians Universität, Geschwister-Scholl-Platz 1, D-80539 München, Germany.

[2] Institut für Experimentelle und Angewandte Physik, Universität Regensburg, D-93040 Regensburg, Germany.

[3] Walter Schottky Institut and Physik Department, Technische Universität München, D-85748 Garching, Germany.

We experimentally investigate the lateral diffusion of dipolar excitons in coupled quantum wells in two (2D) and one (1D) dimensions. In 2D, the exciton expansion obeys non-linear temporal dynamics due to the repulsive dipole pressure at a high exciton density, in accordance with recent reports. In contrast, the observed 1D expansion behaves linearly in time even at high exciton densities. The corresponding 1D diffusion coefficient exceeds the one in 2D by far and depends linearly on the exciton density. We attribute the findings to screening of quantum well disorder by the dipolar excitons.

*e-mail: holleitner@wsi.tum.de

Experiments on exciton traps in quantum well devices aim to observe the bosonic nature of excitons in solid state systems.[1] For detecting the Bose-Einstein condensation of excitons, it is a prerequisite to define confinement potentials for excitons. So far, trapping of excitons has been demonstrated in strained systems,[2]-[4] magnetic traps,[5] "natural traps" defined by interface roughness fluctuations,[6] and electrostatic traps.[7]-[13] As recently reported,[11]-[13] dipolar excitons can be very efficiently trapped in coupled quantum well (QW) heterostructures made of GaAs/AlGaAs with a lithographically structured $SiO_2$-layer on top. There, dipolar excitons are trapped in the plane of the GaAs-QWs just below the perimeter of the $SiO_2$-layers via the electrostatic influence of surface charges at the GaAs/$SiO_2$ interface. Such quasi one-dimensional (1D) channels exhibit a nearly harmonic trapping potential with spring constants of up to 10 keV/cm². Generally, electrostatic traps can be extended towards optoelectronic solid-state devices because of their potential scalability and compatibility with existing semiconductor technology.[14]-[16]

The lateral expansion of excitons has been extensively studied in two dimensions (2D).[13],[15],[17]-[26] At high exciton densities, the interaction of the dipolar excitons leads to a fast, pressure-driven non-linear expansion in 2D.[17],[21],[25] At lower 2D densities, the exciton motion is diffusive, and the corresponding diffusion coefficient has a dependence on the QW width consistent with a universal power law.[17] Here, we demonstrate that the expansion of dipolar excitons in 1D channels obeys a diffusive behavior even at high densities directly after the laser excitation. Surprisingly, the corresponding 1D diffusion coefficient linearly depends on the laser power and therefore, on the exciton density. We observe values of the 1D diffusion coefficient up to 20 times larger than the one found in 2D. Performing equivalent expansion experiments in 2D on the same samples, we again observe the non-linear expansion dynamics as reported in literature.[17],[21],[25] We attribute the findings in 1D to a dynamic screening of the QWs disorder by dipolar excitons at high density. At very low densities, we observe that the excitons are localized in the potential fluctuations along the 1D channel.

Generally, the expansion of a dipolar exciton gas can be described by the following diffusion equation which is extended by a non-linear drift reflecting the dipole-dipole repulsion[25],[26]

$$\frac{\partial n_x}{\partial t} = -\nabla \cdot (\mathbf{J}_{DIFF} + \mathbf{J}_{DD}) - \frac{n_x}{\tau_x} + I, \qquad (1)$$

with $n_x = n_x(x,y,t)$ the exciton density as a function of the in-plane coordinates $x$ and $y$ as well as the time-delay $t$ after the laser excitation, $\tau_x$ the exciton lifetime, $I$ the exciton generation, $\mathbf{J}_{DIFF}(x,y) = -D_x \nabla n_x$ the current density due to diffusion with $D_x$ the diffusion coefficient, and $\mathbf{J}_{DD}(x,y) = -n_x \mu_x e^2 z_0 / \varepsilon_r \cdot \nabla n_x$ the current density due to dipole-dipole repulsion. Here, $z_0$ is the effective out-of-plane spatial separation of the electron and hole wave functions in the QWs defining the excitonic dipole, $\mu_x$ is the exciton mobility, $e$ is the electron charge, and $\varepsilon_r$ is the dielectric constant.[25] A sensitive parameter to estimate the importance of the dipole-dipole interactions in the exciton dynamics is the ratio $\gamma$ between $|\mathbf{J}_{DD}|$ and $|\mathbf{J}_{DIFF}|$, i.e. $\gamma = \gamma(n_X) = |\mathbf{J}_{DD}(n_X)|/|\mathbf{J}_{DIFF}|$.[25] For $\gamma > 1$ directly after the laser excitation, Eq. (1) describes a non-linear expansion driven by repulsive dipole-dipole forces. For $\gamma < 1$ at lower densities, the dipole-dipole interactions are eventually negligible, and Eq. (1) can be solved by the Gaussian distribution

$$n_X(x,y,t) = \frac{N_X}{4\pi D_X t} \exp\left(-\frac{x^2+y^2}{4D_X t}\right), \tag{2}$$

with $N_x$ the number of excitons. The variance $\sigma^2 = 2D_X \cdot t$ of the exciton distribution gives access to the exciton diffusion coefficient and in turn, to the exciton mobility via the Einstein relation (3) $D_X = \mu_X \cdot k_B T_X$, with $k_B$ the Boltzmann constant and $T_X$ the exciton temperature.[15]

Our experiment is performed on an epitaxially grown heterostructure with two 8 nm thick GaAs coupled quantum wells (QWs) separated by a 4 nm $Al_{0.3}Ga_{0.7}As$ barrier [Fig. 1(a)]. The center of the QWs is located 60 nm below the surface of the samples. An n-doped GaAs-layer at a depth of $d$ = 370 nm serves as a back gate, while a semi-transparent titanium-layer is used as the top gate of the field-effect device.[12]-[15] The samples feature an additional $SiO_2$-layer, which is sandwiched between the GaAs surface and the metal top gate. The thickness of the $SiO_2$-layer is ~50 nm, and the titanium top gate has a thickness of ~5 nm. As recently reported,[12],[13] an electrostatic field enhancement in connection with the quantum-confined Stark effect leads to an effective trapping mechanism for dipolar excitons at the perimeter of the $SiO_2$-layer. The trapping potential is indicated by the curved trace in Fig. 1(a). The samples are mounted in a liquid flow cryostat that is positioned under an optical microscope. Using an imaging spectrometer with the entrance slit oriented along the y-direction [Fig. 1(a)], we determine the excitonic recombination energy $E_X$ within the channel at $T_{BATH}$ = 6 K [Fig. 1(b)]. At small y the particular trapping potential can be approximated by a harmonic potential $E_X(y) - E_X(0) = ky^2/2$ with a spring constant $k$ of ~3 keV/cm². For the expansion experiments, we use a mode-locked titanium-sapphire laser with pulses shorter than 150 fs to excite the excitons. The photon wavelength is set to 730 nm, and the time between two successive pulses is tuned to 10 µs by utilizing a pulse picker. On the sample, the laser spot diameter is ~5 µm. At time $t$ after the excitation, the spatial photoluminescence (PL) profile is detected via a fast-gated, intensified charge coupled device camera. An exposure time of 2 ns determines the experiment's time resolution. For a typical PL-image as in Fig. 1(c), we integrate over ~$10^6$ measuring cycles. The triangle in Fig. 1(c) highlights the excitation spot, while one can clearly identify the PL-strip of the excitons being trapped along the perimeter of the $SiO_2$-layer (arrow). The strong capability of our channel to capture excitons allows us to set the excitation spot ~10 µm beside the channel.[13] We thereby avoid heating effects and impurity-PL which are always present at the excitation spot.[27]

We obtain the exciton distribution in the channel along the $x$-direction from line cuts of PL intensity images such as the one in Fig. 1(c). Fig. 2(a) shows such spatially resolved exciton emission profiles along the 1D channel for different $t$. To analyze the dynamics of the expanding excitons quantitatively, we fit the emission profiles by Eq. (2) and extract $\sigma^2$ as a function of $t$ [line in Fig. 2(b)]. For comparison, we move the excitation spot far away from the channel and examine the 2D expansion of the free exciton gas. Then, the radial 2D symmetric PL distribution is projected along the x-direction and fitted according to Eq. (2) [Fig. 2(c)]. The mentioned impurity-PL at the excitation spot gives rise to a small (dashed) center peak on top of the expanding 2D exciton distribution. Noteworthy, Eq. (2) describes a diffusion dominated expansion. As can be seen in Fig. 2(b), Eq. (2) fits the 1D expansion curve remarkably well already at short $t$, while in 2D there are deviations at the tails [arrow in Fig. 2(c)]. In addition, the 1D distribution exhibits several PL-maxima along the x-direction for long $t$ [triangles in Fig. 2(d)]. These peaks are fixed in position, and we do not observe them in 2D. As discussed below, we interpret the PL-maxima to result from excitons localized in potential fluctuations along the 1D channel.

In Fig. 3(a) and (b), the experimentally determined values for $\sigma^2$ are plotted as a function of $t$ for different laser powers $P_{LASER}$ for 2D and 1D. In Fig. 3(a), the continuous lines represent model calculations based upon Eq.(1) with a 2D diffusion coefficient $D_X^{2D} = 14$ cm²/s, $T_X = 6$ K, and an initial $\sigma^2 = 140$ µm² for all laser powers. In the fits, we use following initial exciton densities $n_0^{2D}(x,y,t) = n_X^{2D}(0,0,1\text{ns})$ for the different laser powers: $n_0^{2D}(2.8\ \mu W) = 0.98 \cdot 10^{11}$ cm$^{-2}$, $n_0^{2D}(2.08\ \mu W) = 0.7 \cdot 10^{11}$ cm$^{-2}$, $n_0^{2D}(1.33\ \mu W) = 0.42 \cdot 10^{11}$ cm$^{-2}$, and $n_0^{2D}(0.99\ \mu W) = 0.3 \cdot 10^{11}$ cm$^{-2}$. The ratios of the numerically chosen $n_0^{2D}$ agree within 15 % with the ratio of the corresponding $P_{LASER}$. We find excellent agreement between the calculations and the experimental data [Fig. 3(a)]. Most importantly, for $t > 50$ ns, all curves exhibit the identical linear behavior with the same gradient. Hence, the 2D diffusion coefficient $D_X^{2D}$ can be considered to be independent of $P_{LASER}$. Since only $n_0^{2D}$ is varied to describe the whole set of curves, the initial non-linear expansion for $t < 50$ ns is interpreted to reflect the dipole-dipole repulsion in 2D. Both considerations are in agreement with recent reports.[17],[25]

Fig. 3(b) shows the experimentally determined $\sigma^2$ as a function of $t$ in 1D for the same laser powers as in Fig. 3(a).[27] Strikingly, the 1D expansion is much faster than in 2D. In addition, it exhibits an almost linear increase of $\sigma^2$ already for short $t$. For long $t$, the value of $\sigma^2$ finally saturates, and we detect the PL-maxima as in Fig. 2(d). We note that the saturation value is achieved earlier for lower $P_{LASER}$ [arrows in Fig. 3(b)]. The lines are model calculations based upon a 1D form of Eq.(1) using the same (projected) initial 1D exciton densities as for Fig. 3(a) with $n_0^{1D} = \sqrt{n_0^{2D}(P_{LASER})}$. While there is only a weak non-linear behavior for very small $t \leq 5$ ns in Fig. 3(b), the dominating linear expansion is attributed to a 1D diffusion with respect to Eq.(1). Thus, we introduce an effective power-dependent 1D diffusion coefficient $D_X^{1D} = D_X^{1D}(P_{LASER})$ as a fitting parameter for the whole set of curves in Fig. 3(b). Fig. 4(a) shows the resulting fitting parameter $D_X^{1D}(P_{LASER})$, and that $D_X^{1D}$ depends linearly on $P_{LASER}$. For $P_{LASER} = 2.8$ µW, we determine $D_X^{1D}$ to be 290 cm²/s, which is ~20 times larger than the 2D value $D_X^{2D} = 14$ cm²/s.

Generally, the exciton densities can be deduced from energy resolved PL measurements by a blue-shift of the excitons' recombination energy via (4) $E_{SHIFT} = e^2 z_0/\varepsilon_r \cdot n_X^{2D}$ due to the dipole-dipole repulsion.[25],[26] In the main graph of Fig. 4(b), the resulting $n_X^{2D}$ is plotted as a function of time $t$ for different $P_{LASER}$. As depicted in the inset, $n_X^{2D}(t=1\,\text{ns})$ is directly proportional to $P_{LASER}$, a behavior observed for all time delay $t$. We conclude that in first order, $n_X^{2D}$ is directly proportional to $P_{LASER}$. We repeated such time-dependent PL measurements, while we scanned both perpendicular and along the 1D channels [data not shown]. Since $E_{SHIFT}$ does not vary as a function of the position on the samples, we assume that $n_X^{1D}$ is of same order as $\sqrt{n_X^{2D}}$, and that $n_X^{1D}$ also depends linearly on $P_{LASER}$. Since from Fig. 4(a) we deduced that $D_X^{1D} \propto P_{LASER}$, we finally conclude that $D_X^{1D}$ is directly proportional to $n_X^{1D}$.

Generally, the mobility and hence, the diffusion coefficients [see Eq. (3)] are dominated by scattering at potential fluctuations caused by defects, impurities, as well as interface and alloy fluctuations in the QWs.[17]-[26] As highlighted by triangles in Fig. 2(d), excitons are localized at such potential fluctuations along the 1D channel at small $n_X^{1D}$. Because the dipolar excitons repel each other, we can assume that for larger $n_X^{1D}$, such potential fluctuations are effectively screened by the excitons at the energy bottom of the dipolar exciton gas.[26] Hereby, we interpret the expansion difference in 2D and 1D such that in 1D the exciton diffusion is guided along the 1D channel, along which almost all potential fluctuations are screened by localized excitons at the energy bottom of the exciton gas. The relatively strong confinement of the 1D trapping potential [Fig. 1(b)] will substantially increase the effectiveness of such screening. In addition, it ensures that even for larger $n_X^{1D}$ all excitons still expand along the main direction of the 1D channel in contrast to a free 2D expansion. In this sense, $D_X^{1D}$ defines an upper limit of $D_X^{2D}$, since in 2D there are two linearly independent expansion coordinates. We note that the measured maximum $D_X^{1D}$ corresponds to an elastic mean free path of the trapped excitons of ~1 μm compared to ~50 nm in 2D.[25]

In summary, we present experiments on the lateral expansion of a dipolar exciton gas in 2D and 1D. We find a 2D expansion, which is driven by dipole-dipole interactions. In 1D, the initial expansion is dominated by a linear time-dependence, which we describe as an effective exciton diffusion. Surprisingly, the corresponding diffusion coefficient exceeds the one in 2D and it is directly proportional to the laser power and thus, to the exciton density. We interpret the observations in a way that the potential fluctuations are dynamically screened in 1D.

We thank A. Gärtner, S. Manus and Q. Unterreithmeier for technical support. We gratefully acknowledge financial support by DFG-project KO-416/17-2, the Center for NanoScience (CeNS), and the German excellence initiative via "Nanosystems Initiative Munich (NIM)" and "LMUexcellent".

FIG. 1 (a). Sample sketch. Bent line symbolizes 1D trapping potential below the perimeter of the $SiO_2$-layer on top of a coupled quantum well (QW). Dipolar excitons are excited at the triangle. In all experiments the voltage $V_G$ between the top and back gate is set to -0.2 V. (b) Recombination energy $E_X$ of the excitons as a function of the *y*-position in (a). (c) Spatial image of a typical photoluminescence (PL) distribution. Dashed line indicates the edge of the $SiO_2$-layer.

FIG. 2 (a) and (b). Line cuts of the exciton PL along the 1D channel at $t$ = 4 ns, 13 ns, 31 ns, 55 ns, 91 ns as well as 10 ns with $P_{LASER}$ = 2.08 µW. (c) Line cut in 2D at $t$ = 10 ns. Continuous lines are fits according to Eq. (2). (d) Line cuts in 1D and 2D at $t$ = 180 ns. White triangles highlight PL maxima in 1D. The two graphs are amplified by a factor of 23 compared to the one in (c).

FIG. 3 (a). Variance $\sigma^2$ of the spatial exciton distribution in 2D as a function of *t* at $P_{LASER}$ = 0.99 µW, 1.33 µW, 2.08 µW, and 2.80 µW. (b) Equivalent data along a 1D channel for identical excitation powers. Continuous lines are model calculations w.r.t Eq. (1).

FIG. 4 (a). Diffusion coefficient as a function of $P_{LASER}$ in 1D and 2D (open and closed squares). (b) 2D exciton density as a function of *t* for different $P_{LASER}$. Inset: Initial 2D density at $t \approx$ 1 ns as a function of $P_{LASER}$.

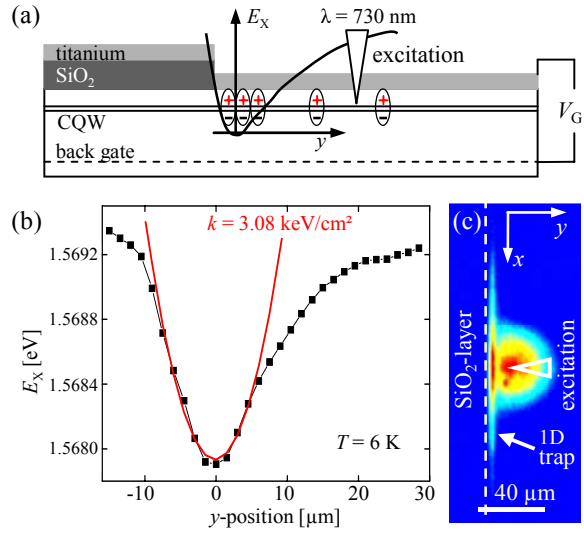

Fig. 1

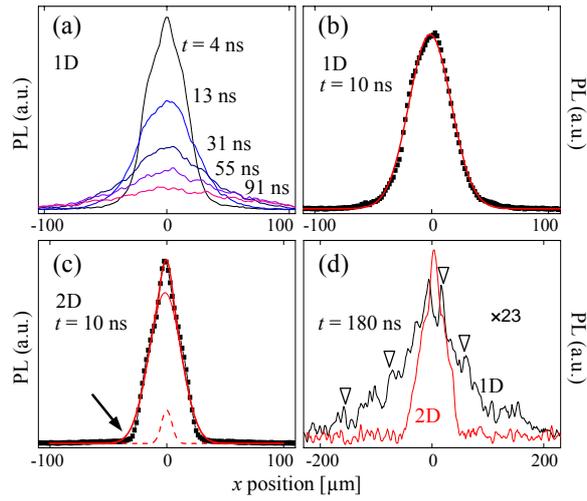

Fig. 2

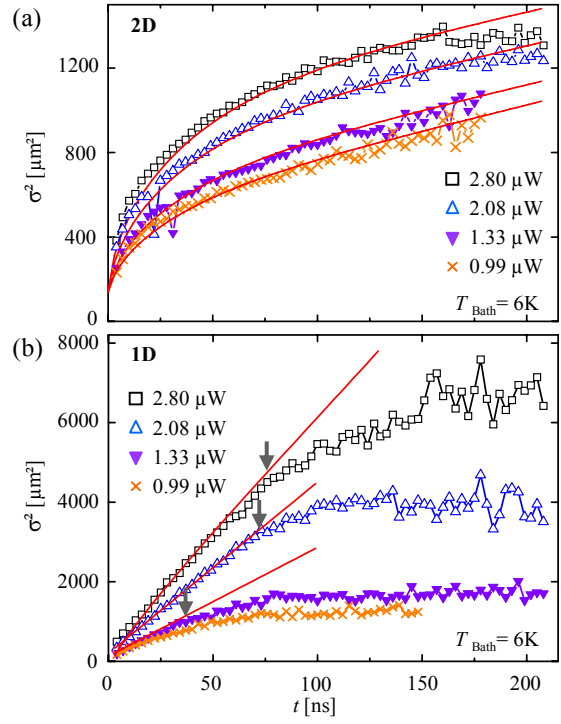

Fig. 3

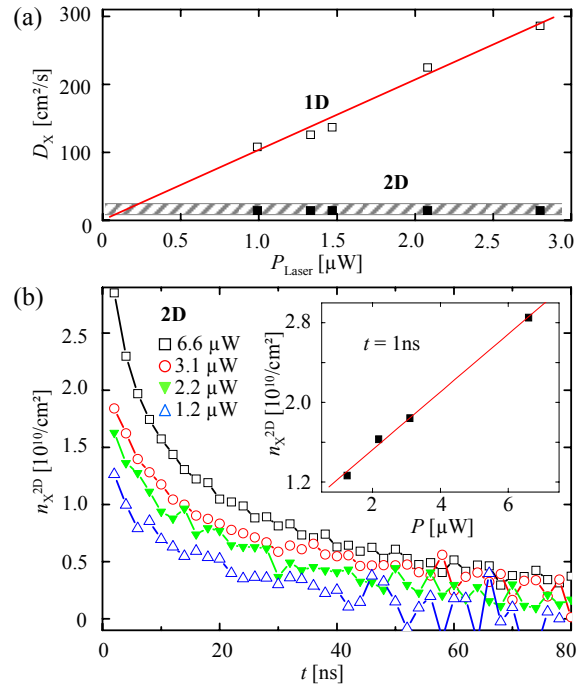

Fig. 4

13